# Pseudo-magnetic field-induced ultra-slow carrier dynamics in periodically strained graphene

Dong-Ho Kang[1†], Hao Sun[1†], Manlin Luo[1†], Kunze Lu[1], Melvina Chen[1], Youngmin Kim[1], Yongduck Jung[1], Xuejiao Gao[1], Samuel Jior Parluhutan[1], Junyu Ge[2], See Wee Koh[2], David Giovanni[3], Tze Chien Sum[3], Qi Jie Wang[1,3], Hong Li[2] and Donguk Nam[1*]

[1]School of Electrical and Electronic Engineering, Nanyang Technological University, 50 Nanyang Avenue, Singapore 639798, Singapore

[2]School of Mechanical and Aerospace Engineering, Nanyang Technological University, 50 Nanyang Avenue, Singapore 639798, Singapore

[3]Division of Physics and Applied Physics, School of Physical and Mathematical Sciences, Nanyang Technological University, 21 Nanyang Link, Singapore 637371, Singapore

[†]These authors contributed equally to this work.

[*]E-mail: dnam@ntu.edu.sg

## Abstract

The creation of pseudo-magnetic fields in strained graphene has emerged as a promising route to allow observing intriguing physical phenomena that would be unattainable with laboratory superconducting magnets. Scanning tunneling spectroscopy experiments have successfully measured the pseudo-Landau levels and proved the existence of pseudo-magnetic fields in various strained graphene systems. These giant pseudo-magnetic fields observed in highly deformed graphene can substantially alter the optical properties of graphene beyond a level that can be feasible with an external magnetic field, but the experimental signatures of the influence of such pseudo-magnetic fields have yet to be unveiled. Here, using time-resolved infrared pump-probe spectroscopy, we provide unambiguous evidence for ultra-slow carrier dynamics enabled by pseudo-magnetic fields in periodically strained graphene. Strong pseudo-magnetic fields of ~100 T created by non-uniform strain in graphene nanopillars are found to significantly decelerate the relaxation processes of hot carriers by more than an order of magnitude. Our finding presents unforeseen opportunities for harnessing the new physics of graphene enabled by pseudo-magnetic fields for optoelectronics and condensed matter physics.

**Introduction**

Since its discovery in 2004[1,2], graphene has continued to revolutionize a wide variety of research fields including physics, electronics and photonics, to name a few[3]. Among the surprising properties is its exceptional mechanical strength[4], which has spurred intense research activity on strain engineering of graphene[5]. Approximately a decade ago, the creation of gauge fields in graphene by harnessing non-uniform strain began attracting considerable attention as a potential route towards realizing previously unattainable physical properties in graphene[6]. Particularly, a theoretical study predicted that a uniquely designed strain in graphene could induce pseudo-magnetic fields that would allow electrons to behave as if they were subjected to a strong real magnetic field[7]. The strength of pseudo-magnetic fields in strained graphene can be orders of magnitude higher than the strength of external magnetic fields generated by superconducting magnets[8–10], thus exhilarating researchers to experimentally verify the existence of such high pseudo-magnetic fields for accessing unprecedentedly high magnetic field regimes. Since then, scanning tunneling spectroscopy (STS) measurements have repeatedly confirmed that pseudo-magnetic fields in mechanically deformed graphene can reach up to a few hundred teslas[11–16]. These giant pseudo-magnetic fields can form significant energy gaps[7] and substantially modify optical transitions in the same manner as an external magnetic field[17–20], but to an unprecedented extent due to the high strength of pseudo-magnetic fields. However, the influence of pseudo-magnetic fields on the optical properties of graphene has not yet been experimentally verified.

In this article, we present an experimental observation of the effect of pseudo-magnetic fields on hot carrier relaxation processes. The strain engineering platform employed in this study allows creating a non-uniform strain distribution with a maximum strain field of ~4% that is a record-high sustainable strain achieved in a deformed graphene sheet thus far. The magnitude of the induced pseudo-magnetic fields is confirmed to be ~100 T via rigorous tight-binding simulations combined with the elasticity theory. Using time-resolved infrared pump-probe spectroscopy combined with theoretical modeling based on many-body interactions, we experimentally confirm that the induced pseudo-magnetic fields in our strained graphene system can decelerate the hot carrier relaxation processes by more than an order of magnitude.

## Results

**Design of strain-engineered graphene nanostructure array**

A typical strain-engineered graphene nanostructure array and the key concept for the generation of pseudo-magnetic fields are illustrated in Figs. 1a and 1b. A monolayer graphene sheet was forced to conform to the topography of the nanopillar array using capillary force (see Methods and Supplementary Note 1 for the detailed fabrication procedure)[21–25]. Graphene's excellent mechanical property[4] allows the accumulation of a large tensile strain at the edges of nanopillars[10]. The resultant non-uniform strain distribution near the edges of nanopillars can generate strong pseudo-magnetic fields, which force electrons to move in a circular motion as if they are under a strong external magnetic field[7,8]. Due to the preserved global time-reversal symmetry under the influence of strain-induced pseudo-magnetic fields, electrons in the K and K′ valleys experience pseudo-magnetic fields of opposite signs, thus circulating in opposite directions[12]. This cyclotron motion of the charge carriers corresponds to the creation of pseudo-Landau levels (Fig. 1b), which enable us to observe substantially modified optical transitions in a strained graphene device. Figures 1c and 1d show scanning electron microscopy (SEM) and atomic force microscopy (AFM) images of the strained graphene device used in this study. The structural analyses reveal that graphene is strongly deformed at the sharp corners and edges without showing any signature of fracture across the entire array.

**Formation of strong pseudo-magnetic fields in highly strained graphene**

Figure 2a presents the Raman spectrum measured for strained graphene on the nanopillar (red) (see Methods for more details on Raman spectroscopy); for comparison, the Raman spectrum for the unstrained control device (black) is also presented. Two-dimensional Raman mapping was performed to confirm the highly periodic nature of our strained nanopillar array (Fig. 2a, inset). The measured 2D Raman peak shift of ~85.2 cm$^{-1}$ is a record-high value achieved thus far for deformed graphene on nanostructured substrates[22–25], demonstrating the excellence of our strain engineering platform (see Supplementary Note 2 for more details on G and 2D peak shift analyses). Figure 2b shows the local strain distribution based on the experimentally obtained topographic information (see Supplementary Note 3 for the strain calculation). To calculate the local strain distribution, we employed a reconstructed 3 × 1 strained graphene nanopillar array (top panel), in which the brightest and the darkest regions correspond to the 300-nm and 0-nm heights of the structure, respectively. Strong structural deformation near the sharp corners and edges of the nanopillars generated a substantial tensile strain of up to ~3.5% at the atomic scale. By decomposing the strain distribution along the x and y directions as shown in the middle ($\epsilon_{xx}$) and bottom ($\epsilon_{yy}$) panels, respectively, we determined that $\epsilon_{yy}$ is nearly absent where $\epsilon_{xx}$ is maximum, indicating that our strained graphene nanopillar is largely under uniaxial strain. A maximum experimental strain value of 1.3% was derived from the measured Raman shift (Fig. 2a) by using a strain-shift coefficient[26] (see Methods). This discrepancy between the simulated and measured maximum strain values is attributable to the inherent resolution limit of the Raman measurement with the diffraction-limited laser spot size[21]. By performing convolution on the simulated atomic-scale strain distribution by using a two-dimensional Gaussian corresponding to the spot size of the laser beam[21], a realistic strain distribution that could be optically measured on the nanopillar was deduced, yielding a maximum strain value of 1.32% (Fig. 2c). A comparison between the measured strain from the Raman shift and the convoluted strain obtained from atomic-scale simulations revealed excellent quantitative agreement. The spatial distribution of the pseudo-magnetic fields ($B_S$) at the atomic scale (Fig. 2d) was obtained using a well-developed method based on the tight-binding simulation[6–9,11,12,27], which couples the Dirac equation to the deformed graphene surface to obtain the following relationship between strain fields and gauge fields:

$$A_x = \frac{\beta}{2a_0}(\epsilon_{xx} - \epsilon_{yy}), \quad A_y = \frac{\beta}{2a_0}(-2\epsilon_{xy}),$$

where $\beta$ is a constant (~3), $a_0$ is the nearest carbon–carbon bond length (~0.14 nm), and $\epsilon$ is a 2 × 2 strain tensor. This simulation method has successfully replicated the atomic-scale experimental pseudo-magnetic field distribution measured by STS[11,12,16], thus ensuring the reliability of our simulation (see Supplementary Note 4 for more details on the pseudo-magnetic field simulation). As shown in Fig. 2d, the pseudo-magnetic fields reach up to approximately 100 T near the sharp edges and corners that host the largest deformation and the steepest strain gradient (Fig. 2b), which is largely consistent with ref. 10.

**Pseudo-magnetic field-induced ultra-slow carrier dynamics**

To study the influence of pseudo-magnetic fields on the optical properties of graphene, we performed ultrafast pump-probe spectroscopy (see Methods for measurement details). Figure 3a presents a conceptual illustration of our femtosecond pump-probe measurement setup. The devices were pumped by femtosecond laser pulses while time-delayed probe pulses were used to monitor the photo-induced change in the reflection spectrum ($\Delta R/R$), which is directly proportional to the absorption in graphene for monolayer devices[28–30]. The measurement was performed at 4 K, unless stated otherwise, and low pump fluence (<1 mJ cm$^{-2}$) was used to avoid any heating effect (see Methods for details). The probe pulse spot size was approximately $10 \times 10$ μm$^2$, thus allowing us to probe a large number (>16) of highly identical nanopillars simultaneously for a large effective area possessing sizable pseudo-magnetic fields. The strength of pseudo-magnetic fields is negligible in graphene placed on flat surfaces (Fig. 2d) where there is little out-of-plane lattice distortion (Fig. 2b), which is predominantly responsible for building up pseudo-magnetic fields[12]. Within the area of investigation governed by the probe beam size, therefore, pristine graphene with massless Dirac cones (Fig. 3b, left) existed simultaneously with strained graphene in which pseudo-Landau levels were attained (Fig. 3b, right), and both contributed to the measured data presented in Figs. 3c and 3e.

To better understand the carrier dynamics of the device on nanopillars possessing pseudo-magnetic fields, we first performed control experiments involving the same monolayer graphene sheet on an entirely flat surface without nanopillars. In pristine graphene with massless Dirac cones (Fig. 3b, left), the optically excited carriers by pump pulses ($\lambda_{pump}$~1,030 nm) relax to the lower energy states extremely rapidly within a few tens of femtoseconds through electron–electron scattering, thus inhibiting the absorption of probe pulses at the corresponding energy ($\lambda_{probe}$~1,450 nm) due to Pauli blocking[30-43]. This decreased absorption in the control device was reflected in the rapid rise of the negative reflection change ($-\Delta R/R$) (Fig. 3c, black empty circles). Subsequently, the number of charge carriers in the particular energy state rapidly reduces within 100–200 fs through electron–electron and electron–optical phonon scattering processes[30–35], and this was revealed by the rapid decay of $-\Delta R/R$. The evolution of these extremely fast relaxation processes occurred within the temporal resolution of our probe pulse (~230 fs). A Subsequent slow relaxation process with a very weak intensity via the emission of acoustic phonons[36] can also be observed in our sample, but only at higher pump fluences. In this study, we focused on the initial rapid carrier relaxation processes because of their importance in various crucial device applications[19,43–49].

The response of the nanopillar device is also presented in the same figure (Fig. 3c, red empty circles). Immediately after pump excitation, $-\Delta R/R$ of the nanopillar device quickly rose and decayed on the same time scale as that of the control device. Surprisingly, $-\Delta R/R$ rose again very slowly until ~700 fs, followed by a slow decay with a time constant of ~1.66 ps. To illustrate the two distinct regimes, a detailed view over a shorter time range is shown in the inset of Fig. 3c. To understand the origins of these regimes, we turn to Fig. 3b. As mentioned earlier, both pristine graphene and strained graphene with pseudo-magnetic fields contributed to the signal from the nanopillar device. The rapid change in $-\Delta R/R$ in the rapid regime can be clearly attributed to pristine graphene with fast carrier relaxation processes (Fig. 3b, left). Unlike pristine graphene, the relaxation process in strained graphene can be markedly influenced by pseudo-Landau levels (Fig. 3b, right). First, the initial electron–electron scattering of the optically excited carriers in higher energy states corresponding to the pump energy can be suppressed in the presence of pseudo-Landau levels, which substantially slow down the filling process in the lower energy states corresponding to the probe energy. The pseudo-Landau levels also impede subsequent electron–electron and electron–optical phonon scattering processes, resulting in an ultra-slow depletion of hot carriers from the probe energy states. The hypothesized carrier dynamics of the strained graphene with pseudo-Landau levels is unambiguously captured in the temporal evolution of $-\Delta R/R$ in the slow regime. This ultra-slow carrier dynamics in the presence of pseudo-Landau levels can be explained by the significant reduction of phase space for scattering processes, which has also been experimentally observed in pristine graphene[50] and other materials[18,51], but only in the presence of an external magnetic field.

To further support the role of pseudo-Landau levels in producing these effects, we investigated the temperature-dependent carrier dynamics of both the control device and the nanopillar devices. Although there was no appreciable change in the carrier dynamics at different lattice temperatures in the control device (Fig. 3d), the nanopillar device exhibited substantially slower carrier dynamics at 4 K with respect to that at 300 K (Fig. 3e). Previous studies have confirmed that the initial fast relaxation dynamics of pristine graphene is insensitive to the lattice temperature due to the extremely efficient electron–electron and electron–optical phonon processes[30,32,50,52]. In contrast, a strong temperature dependence of the carrier dynamics was observed in Landau-quantized graphene with an external magnetic field, and this was attributed to the significantly reduced electron–electron scattering at low temperatures in the presence of Landau quantization[50]. The contrasting temperature dependence of the carrier dynamics, as shown in Figs. 3d and 3e, is highly consistent with the previously observed contrasting phenomena in pristine graphene and Landau-quantized graphene, thus providing another strong evidence that points to the formation of pseudo-Landau levels in the nanopillar device. We also observed that the decay becomes slower at lower pump fluences due to the reduced electron–electron scattering (Supplementary Fig. 5).

**Theoretical modeling of carrier dynamics under the influence of pseudo-magnetic fields**
Optical Bloch equations based on many-body interactions were used to quantitatively understand the relaxation dynamics of charge carriers under the influence of pseudo-magnetic fields. The main many-body interaction process considered in the modeling was the electron–electron scattering that is the dominant relaxation mechanism in Landau-quantized graphene[50]. We model the dynamics of the charge carriers for the initial excitation to the pump energy states and the filling-in to the probe energy states as illustrated in Fig. 3b. The subsequent depletion of the charge carriers out of the probe energy states was also considered (see supplementary Note 6 for the details on theoretical modeling). Figure 4 presents the calculated electron population in the probe energy states as a function of the delay time. The electron population is directly proportional to the decreased absorption and the negative reflection change ($-\Delta R/R$) (Figs. 3c–e) because the occupied probe energy states suppress the absorption of probe pulses due to Pauli blocking as explained earlier. The rise time could be significantly increased up to 800 fs at a pseudo-magnetic field intensity of 80 T. This result reveals a clear slow-down of the carrier relaxation processes under the influence of strong pseudo-magnetic fields, which is in reasonable agreement with our experimental results as shown in Figs. 3c and 3e. The discrepancy between the theoretical and experimental results may be attributed to the non-uniform pseudo-magnetic field distribution.

## Discussion

In summary, we have presented an experimental observation of the effect of pseudo-magnetic fields on the carrier relaxation processes in highly strained graphene. Our periodic nanopillar structure array achieves strong pseudo-magnetic fields of up to 100 T, which play a major role in enabling the observation of an extended hot carrier relaxation lifetime by more than an order of magnitude. The pseudo-magnetic field-induced ultra-slow carrier dynamics in highly strained graphene poses crucial implications towards the realization of a new class of graphene-based optoelectronic devices, including pseudo-Landau level lasers[44] and highly efficient hot electron harvesting devices[43]. The light emission[45,53] and population inversion properties[46] in graphene can be improved drastically by slowing down the hot carrier relaxation processes through pseudo-magnetic fields. Our strain engineering platform can also be used to enable pseudo-magnetic field-based valleytronics[48,49]. By presenting experimental evidence of the effect of pseudo-magnetic fields, our finding offers a new landscape of opportunities towards creating pseudo-magnetic field-based graphene optoelectronic devices.

## Methods

**Device fabrication.** A nanostructured substrate was fabricated by buffered oxide etch (BOE)-based wet-etching process. For an etching mask (an array with 1-μm diameter holes formed at 1.6-μm intervals), we patterned polymethyl methacrylate resist (950 PMMA A6, MICROCHEM) on SiO$_2$/Si substrate using Raith e-line e-beam lithography system built in field-emission SEM (JEOL JSM-7600F). The substrate was then dipped in BOE (12.5% HF, 87.5% NH$_4$F) with 3 min 30 sec at room temperature, followed by atomic layer deposition (ALD) for depositing a 20-nm Al$_2$O$_3$ layer on the entire substrate. Graphene monolayer was transferred onto the nanostructured substrate via wet-transfer technique. To make a tight adhesion between graphene and nanostructures, capillary force-induced drying technique was employed[21]. Further details on the fabrication process are provided in Supplementary Note 1.

**Structural characterization and Raman spectroscopy.** To analyze the structural characteristics and strain distribution of our nanopillar devices, we used a field-emission SEM (JSM-7600F, JEOL), AFM (Park XE 15, Park systems), and Raman spectroscopy (Alpha300 M+, WITec). The electron acceleration voltage in SEM was set to 10 kV. AFM was performed on nanopillar sample with tapping mode. The scan size, resolution and speed were set to $10 \times 10$ μm$^2$, $512 \times 512$ pixel$^2$ and 0.5 Hz, respectively. For Raman spectroscopy, a 532-nm excitation laser was used along with a 100× objective lens. During the measurement, additional care was taken to keep the laser power low enough to avoid any heating effect. The strain-shift coefficient of 65.4 cm$^{-1}$/%[26] was used to estimate the maximum strain of 1.3% in our nanopillar device.

**Femtosecond pump-probe measurement**. Femtosecond pump-probe measurement was performed in a reflection geometry. The sample was held in a closed-cycle helium cryostat (Cryostation s50, Montana Instrument). The pump pulses were generated from a femtosecond laser with a 230-fs pulse duration (CARBIDE, Light Conversion). The differential reflectance change was measured with a probe pulse ($\lambda_{probe}$ = 1450 nm) centered at the same spatial position. The spot size of pump and probe pulses were approximately 100 × 100 μm$^2$ and 10 × 10 μm$^2$, respectively. A pump fluence was approximately 0.4–1 mJ cm$^{-2}$, which was adjusted by a continuously variable neutral density filter.

53. Kim, Y. D. *et al.* Bright visible light emission from graphene. *Nat. Nanotechnol.* **10**, 676–681 (2015).

**Acknowledgements**

The research of the project was in part supported by Ministry of Education, Singapore, under grant AcRF TIER 1 2019-T1-002-050 (RG 148/19 (S)). The research of the project was also supported by Ministry of Education, Singapore, under grant AcRF TIER 2 (MOE2018-T2-2-011 (S)). This work is also supported by National Research Foundation of Singapore through the Competitive Research Program (NRF-CRP19-2017-01). This work is also supported by National Research Foundation of Singapore through the NRF-ANR Joint Grant (NRF2018-NRF-ANR009 TIGER). This work is also supported by the iGrant of Singapore A*STAR AME IRG (A2083c0053). T.C.S and D.G. acknowledge the financial support from Nanyang Technological University under the start-up grants M4080514 and M4081630. Q.J.W. acknowledge the financial support from Singapore Ministry of Education Academic Research Fund Tier 2 under grant no. MOE2018-T2-1-176. The authors would like to acknowledge and thank the Nanyang NanoFabrication Centre (N2FC).
**Author contributions**

D.-H.K., H.S., M.L. and D.N. conceived the initial idea of the project. Under the guidance of Q.J.W., H.L. and D.N., D.-H.K., M.L., X.G. and S.P. fabricated the samples. D.-H.K., M.L., K.L. and X.G. performed the Raman measurements and analysis. H.S. performed the simulation and modelling. D.-H.K., H.S. and M.C. carried out the pump-probe measurements. Under the guidance of T.C.S. and D.N., D.-H.K., D.G. and H.S. performed data analysis. Under the guidance of H.L., J.G. and S.W.K. performed AFM measurements. Y.J. and Y.K. helped low-temperature measurements. D.-H.K., H.S., M.L., K.L. and D.N. wrote the manuscript. All authors revised the manuscript. D.N. supervised the entire project.

## Additional information

Supplementary information is available in the online version of the paper. Correspondence and requests for materials should be addressed to D.N.

## Competing financial interests

The authors declare no competing financial interests.

# Figures

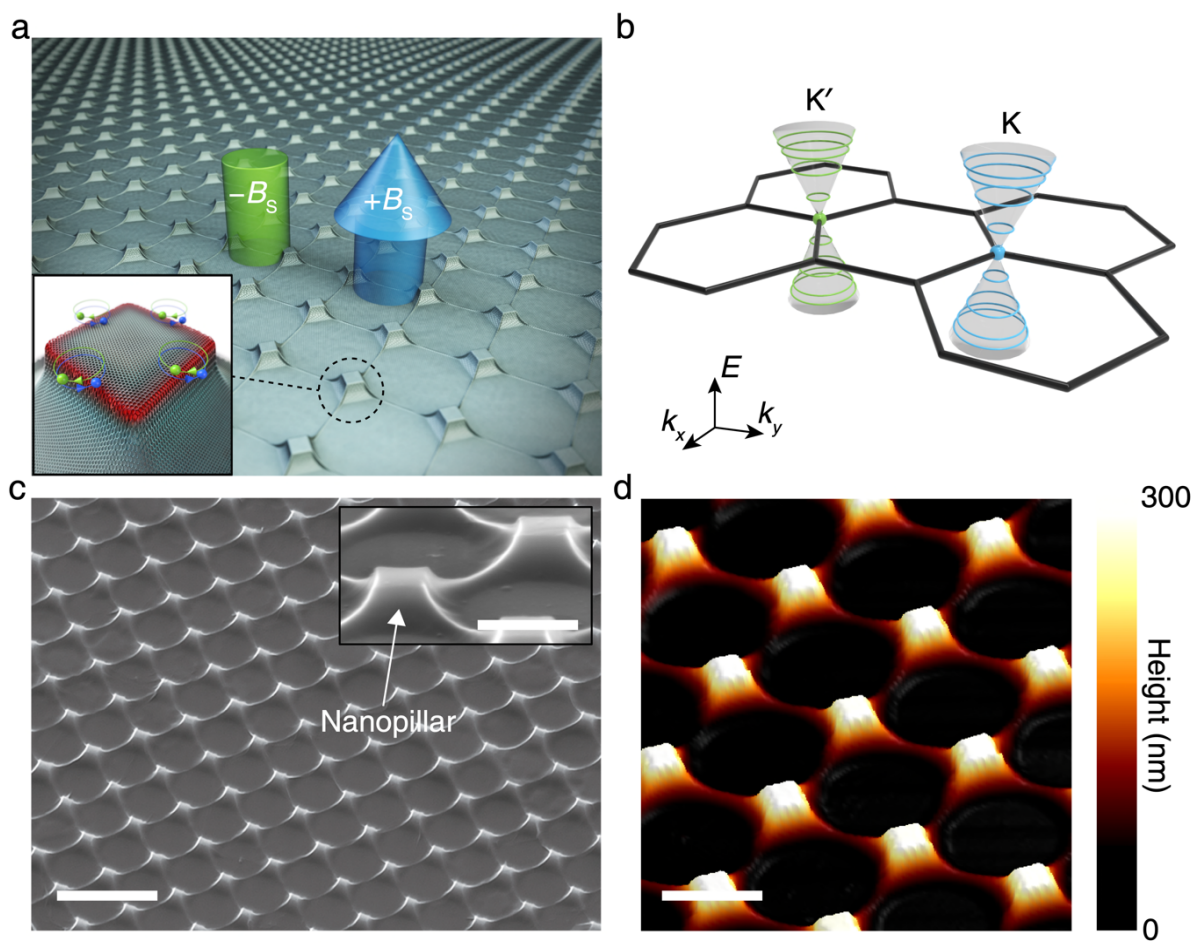

**Figure 1 | Design of a strain-engineered graphene nanostructure array. a,** Schematic illustration of a strained graphene nanopillar array possessing pseudo-magnetic fields of opposite signs. Inset: magnified view of a single strained graphene nanopillar structure. Non-uniform tensile strain at the edges of the nanopillar induces pseudo-magnetic fields of opposite signs ($-B_S$ and $+B_S$) in the two valleys of the graphene band structure, thus forcing electrons in the K and K′ valleys to circulate in the opposite directions. **b,** Schematic illustration showing Landau quantization at the K and K′ points in the first Brillouin zone of graphene in the presence of pseudo-magnetic fields. **c,** Tilted-view SEM image of a strained graphene nanopillar array. Scale bar, 2 μm. Inset: magnified SEM image. Scale bar, 500 nm. **d,** AFM topography of the strained graphene nanopillar array. Scale bar, 1 μm.

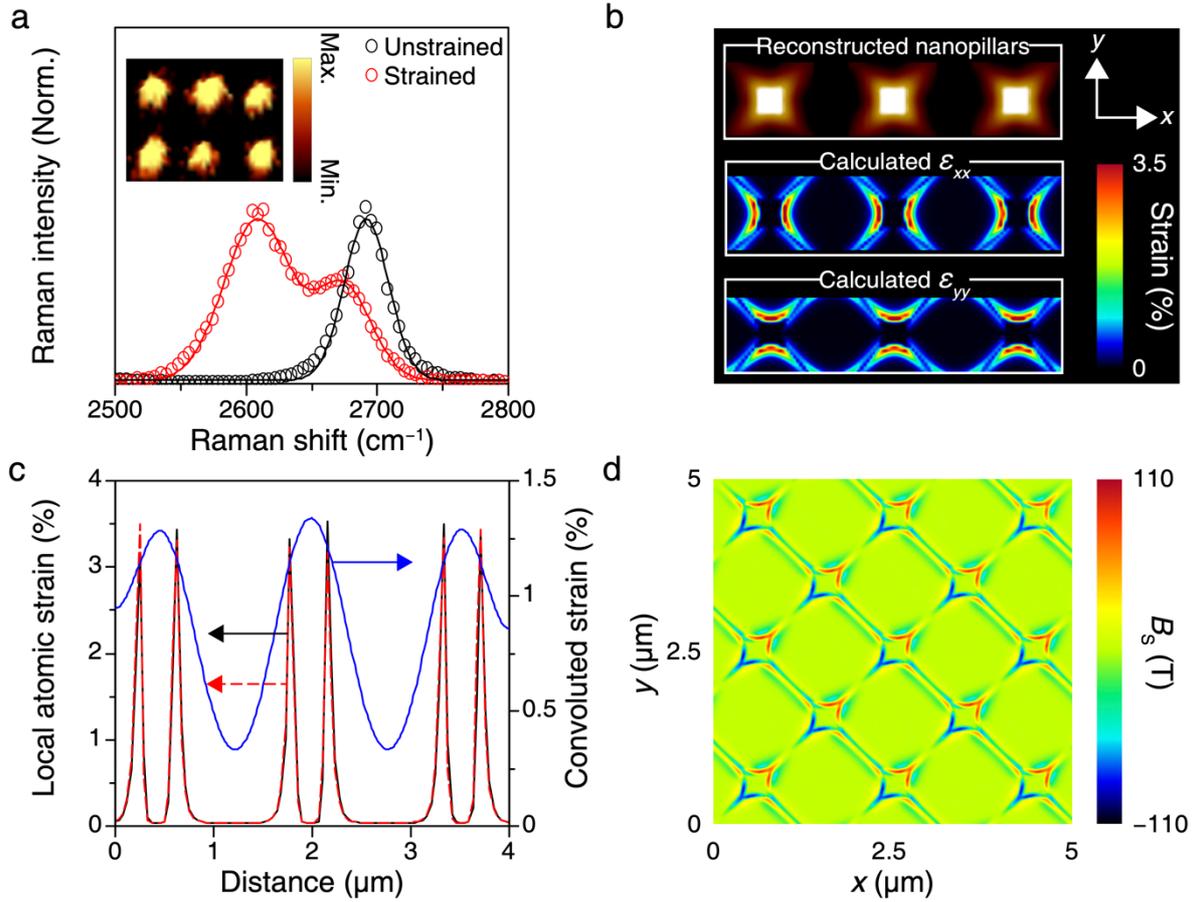

**Figure 2 | Formation of strong pseudo-magnetic fields in highly strained graphene. a,** Raman spectra of unstrained (black) and strained (red) graphene. Symbols are measurement data; curves are fitting data. Inset: Two-dimensional Raman mapping data plotting the 2D peak frequency of a 3 × 2 strained graphene nanopillar array. **b,** The topographic image of the reconstructed a 3 × 1 strained graphene nanopillar array (top panel) and corresponding local strain distributions along the x ($\epsilon_{xx}$, middle panel) and y ($\epsilon_{yy}$, bottom panel) directions. The brightest and darkest regions in the topographic image indicate 300-nm and 0-nm heights of structure, respectively. **c,** Cross-section of the strain profile showing local atomic strain for $\epsilon_{xx}$ (black solid line) and $\epsilon_{yy}$ (red dashed line), and convoluted strain distribution (blue solid line). **d,** Pseudo-magnetic fields distribution in strained graphene nanopillar array.

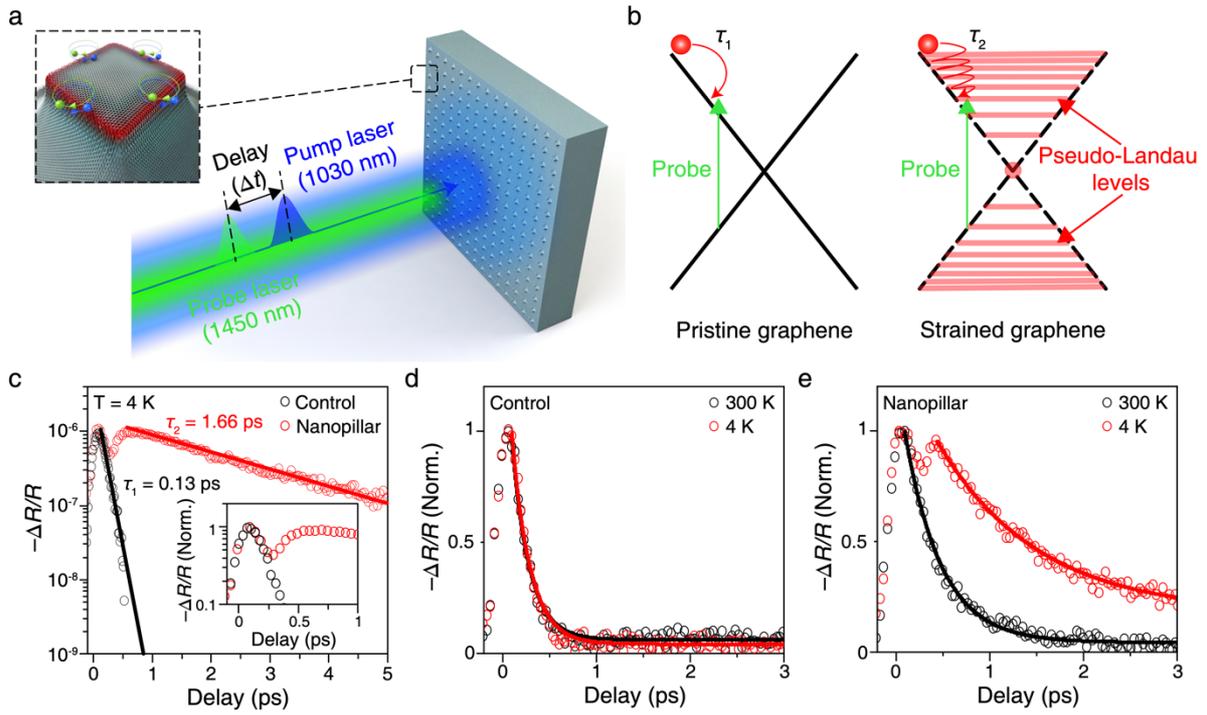

**Figure 3 | Pseudo-magnetic field-induced ultra-slow carrier dynamics. a**, Schematic illustration of our femtosecond pump-probe measurement setup. Pump pulses (blue) excite charge carriers and probe pulses (green) are used to monitor the sample responses at different delay times (Δ$t$) after the pump pulses arrive at the sample. Left inset: magnified image of a single nanopillar structure. **b**, Schematic illustration of relaxation process of photoexcited carriers in pristine graphene with massless Dirac cones (left) and strained graphene attaining pseudo-Landau levels (right). The formation of pseudo-Landau levels can significantly decelerate the relaxation process, resulting in a longer decay time in strained graphene ($\tau_1 \ll \tau_2$). **c**, Measured reflection change as a function of the delay time on control (black) and nanopillar (red) samples. Symbols are measurement data; lines are fitting data for the decay region. The decay times of both samples are extracted from fitting data; $\tau_1$ = 0.13 ps (Control) and $\tau_2$ = 1.66 ps (Nanopillar). Inset: Normalized reflection change of both samples from –0.15 to 1 ps. **d and e**, Normalized reflection change of control (d) and nanopillar (e) samples measured at 300 K (black) and 4 K (red). Symbols are measurement data; lines are fitting data for the decay region.

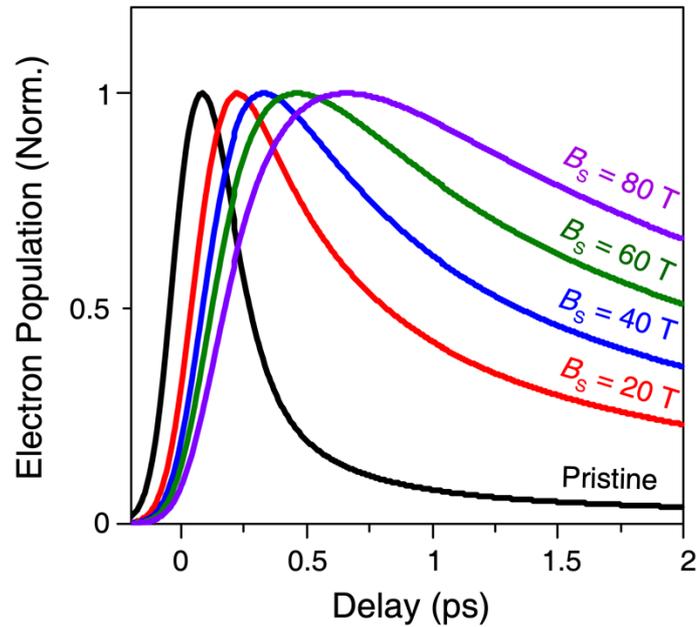

**Figure 4 | Theoretical modeling of carrier dynamics under the influence of pseudo-magnetic fields.** Calculated electron population in the probe energy states as a function of the delay time for strained graphene with different pseudo-magnetic fields ($B_S$) of 0 T (black), 20 T (red), 40 T (blue), 60 T (green), and 80 T (purple). All graphs are normalized. With the increased pseudo-magnetic field intensity, the rise time increases significantly up to 800 fs for $B_S$ = 80 T.

# Supplementary Information

## Pseudo-magnetic field-induced ultra-slow carrier dynamics in periodically strained graphene

Dong-Ho Kang[1†], Hao Sun[1†], Manlin Luo[1†], Kunze Lu[1], Melvina Chen[1], Youngmin Kim[1], Yongduck Jung[1], Xuejiao Gao[1], Samuel Jior Parluhutan[1], Junyu Ge[2], See Wee Koh[2], David Giovanni[3], Tze Chien Sum[3], Qi Jie Wang[1,3], Hong Li[2] and Donguk Nam[1*]

[1]School of Electrical and Electronic Engineering, Nanyang Technological University, 50 Nanyang Avenue, Singapore 639798, Singapore
[2]School of Mechanical and Aerospace Engineering, Nanyang Technological University, 50 Nanyang Avenue, Singapore 639798, Singapore
[3]Division of Physics and Applied Physics, School of Physical and Mathematical Sciences, Nanyang Technological University, 21 Nanyang Link, Singapore 637371, Singapore

[†]These authors contributed equally to this work.
[*]E-mail: dnam@ntu.edu.sg

## Table of Content

### Note 1. Fabrication procedure

*Fabrication of a nanostructured substrate:* Supplementary Figure 1 illustrates the fabrication procedure for a nanostructured substrate. We first spin-coated a polymethyl methacrylate (PMMA) resist (950 PMMA A6, MICROCHEM) on a 300-nm-thick $SiO_2$/Si substrate at 4500 rpm for 30 sec, followed by baking at 180 ºC for 2 min (Supplementary Figs. 1a-b). Then, an etching mask (an array with 1-µm diameter holes formed at 1.6-µm intervals) was patterned (Supplementary Fig. 1c) using Raith e-line e-beam lithography system. The substrate was then immersed in buffered oxide etch (BOE) (12.5% HF, 87.5% $NH_4F$) with 3 min 30 sec at room temperature (Supplementary Fig. 1d). PMMA resist was removed by acetone (80 ºC, 10 min), isopropyl alcohol (IPA) (room temperature, 2 min), deionized (DI) water (room temperature, 2 min), and $O_2$ plasma treatment (150 W, 3 min), followed by atomic layer deposition (ALD) to deposit a 20-nm $Al_2O_3$ layer on the entire substrate (Supplementary Figs. 1e-f).

*Graphene wet transfer process:* Supplementary Figure 2 shows the detailed schematic illustration for graphene wet transfer process. A PMMA supporting layer (950 PMMA A6, MICROCHEM) was spin-coated onto a graphene/$SiO_2$/Si substrate at 4500 rpm for 90 sec, followed by baking at 180 ºC for 2 min (Supplementary Figs. 2a-b). A high-quality large-area chemical vapor deposition (CVD)-grown monolayer graphene was used in our study. The PMMA-covered graphene layer was floated by 5% diluted hydrofluoric acid (HF) solution by etching away the underlying $SiO_2$ layer. After the PMMA/graphene layer was lifted off in the HF solution, it was transferred to DI water to remove HF residue. The nanostructured substrate was then used to fish the PMMA/graphene layer (Supplementary Fig. 2c). The sample was then dried at room temperature while standing at an angle to make a tight adhesion between graphene and the nanostructured substrate by capillary force[1] (Supplementary Fig. 2d). Afterwards, the PMMA resist was removed by acetone, IPA, and DI water (Supplementary Fig. 2e).

**Note 2. Raman analysis on unstrained and strained graphene**

The Raman spectra of unstrained and strained graphene fabricated using the previously described procedure are both plotted in Supplementary Fig. 3. Supplementary Figure 3a compares the Raman spectra of the unstrained and strained graphene near the G mode peak, which arises from the doubly degenerate zone center phonons at $\Gamma$[2]. Similar to the 2D peak, tensile strain in the nanopillar causes the G peak to red-shift away from the unstrained graphene peak at 1581.9 cm$^{-1}$. Furthermore, strain breaks the rotation symmetry of the zone center phonons[3], creating G-peak splitting with two peaks at 1544.87 cm$^{-1}$ and 1574.7 cm$^{-1}$. The 2D peak shown in Supplementary Fig. 3b, arising from the second order scattering of the zone-boundary phonons, has a peak value of 2691.7 cm$^{-1}$ in unstrained graphene. By applying tensile strain, the graphene C-C bonds are elongated and become weaker, causing the 2D peak to red-shift. The splitting of the two peaks into 2606.5 cm$^{-1}$ and 2674.3 cm$^{-1}$ shows that the C-C bonds are being stretched by varying degrees[4].

## Note 3. Calculation of local strain distribution for strained graphene nanopillars

Structural distortion-induced strain in graphene nanopillars can be evaluated using the strain tensors, which are characterized as[5]:

$$\epsilon_{ij}(r) = \frac{1}{2}\left(\partial_i u_j(r) + \partial_j u_i(r) + \partial_i h(r)\partial_j h(r)\right), \quad (1)$$

where $u(r)$ and $h(r)$ represent the in-plane and out-of-plane deformation fields, respectively. We mainly focus on the out-of-plane deformation induced by nanopillars, since the resulting in-plane lattice distortion is much smaller than the vertical lattice distortion[6]. Consequently, from given atomic structural distortions of graphene, which can be determined by high-resolution atomic force microscopy (AFM), the strain components can be calculated as:

$$\epsilon_{xx} = \frac{1}{2}\left(\frac{\partial h(r)}{\partial x}\right)^2, \quad (2)$$

$$\epsilon_{yy} = \frac{1}{2}\left(\frac{\partial h(r)}{\partial y}\right)^2, \quad (3)$$

$$\epsilon_{xy} = \frac{1}{2}\frac{\partial h(r)}{\partial x}\frac{\partial h(r)}{\partial y}, \quad (4)$$

where $h(r)$ is the AFM topographic data, and $r$ denotes the position in the x-y plane. The resulting strain distributions are presented in Fig. 2c, which are also used to calculate the spatial distribution of pseudo-magnetic fields as explained in Supplementary Note 4.

**Note 4. Calculation of pseudo-magnetic fields for strained graphene nanopillars**

The Hamiltonian modified by pseudo-gauge fields induced by strain in wave-vector space can be written as[7]:

$$H(q) = v_0 \vec{\sigma} \cdot (\vec{q} + \vec{A}), \quad (5)$$

where $v_0 = \frac{3a_0 t_0}{2}$ is the Fermi velocity, and the typical value is about $10^6$ ms$^{-1}$, $a_0 = 0.14\ nm$ is the length of the carbon-carbon bond, $t_0 = 2.7$ is the hopping amplitude in the tight-binding model. $\vec{\sigma}$ is the 2 × 2 Pauli matrix for the sublattice freedom. The pseudo-gauge field or vector potential $\vec{A}$ arising from the strain effect is expressed as[5,7]:

$$A_1 = \frac{\beta}{2a_0}(\epsilon_{11} - \epsilon_{22}), \quad A_2 = \frac{\beta}{2a_0}(-2\epsilon_{12}), \quad (6)$$

Where $\beta \approx 3$ is a constant. Since strain does not break the time reversal symmetry, the vector potential $\vec{A}$ will possess opposite signs for the two inequivalent $K$ and $K'$ points, making the net magnetic field to be zero. The pseudo-magnetic field perpendicular to the surface is given by $\vec{B} = \nabla \times \vec{A}$ and can be written as a function of the strain tensors[7]:

$$B_p = \frac{-\beta}{2a_0}\left(\partial_y \epsilon_{xx}(x,y) - \partial_y \epsilon_{yy}(x,y) + 2\partial_x \epsilon_{xy}(x,y)\right). \quad (7)$$

One can see that the pseudo-magnetic field, $B_p$, vanishes for a uniform strain and that only non-uniform strain contributes to the finite value of the induced field strength. When the pseudo-magnetic field is relatively uniform under a specific strain field (e.g., a well-designed triaxial strain), the Dirac bands can be rearranged into non-equidistant pseudo-Landau levels having energies of $E_n$, which can be approximated as:

$$E_n = sgn(n)\hbar v_0 \sqrt{\frac{2e_0}{\hbar}|nB_p|}, \quad (8)$$

where $n$ is the Landau level index and $e_0$ is the electron charge. One can insert the values of all constants, and obtain a simplified one:

$$E_n = 36 sgn(n)\sqrt{|nB_p|}\ (\text{meV}). \quad (9)$$

**Note 5. Calculation of local density of states (LDOS) of strained graphene nanopillars**

LDOS with different pseudo-magnetic fields can be calculated by the following expression:

$$DOS(E) = \frac{1}{\pi} \sum_n \frac{\gamma}{(E-E_n)^2 + \gamma^2}, \quad (10)$$

where $\gamma$ is the broadening factor of Landau level. Supplementary Figure 4 shows the calculated LDOS for the strained graphene with different pseudo-magnetic fields (10, 40, and 80 tesla).

**Note 6. Theoretical modeling for the pseudo-magnetic field effect on carrier dynamics**

*6.1 Classic light-matter interaction and pumping pulse*

We consider a classic light-carrier coupling for pumping process. The excitation pulse can be written as[8]:

$$\vec{A}(t) = A_{env} \left( A_0^+ \begin{pmatrix} \cos(\omega t) \\ \sin(\omega t) \end{pmatrix} + A_0^- \begin{pmatrix} \cos(\omega t) \\ -\sin(\omega t) \end{pmatrix} \right), \quad (11)$$

where $A_{env}$ is the envelope function that has the following form:

$$A_{env} = \frac{1}{\omega} \sqrt{\frac{2\sqrt{\ln 2} e_{pf}}{\sqrt{\pi} \epsilon_0 c \sigma_{FWHM}}} e^{-\frac{2\ln 2 t^2}{\sigma_{FWHM}^2}}, \quad (12)$$

where $e_{pf}$ is the pump fluence, $\omega$ is the frequency of pump light, $\epsilon_0$ is the dielectric constant, and $\sigma_{FWHM}$ is the full width at half maximum (FWHM) of the pump light.

*6.2 Quantum mechanical description of many-particle system*

The full many-particle Hamiltonian with electron, phonon and photon parts in our system can be expressed as:

$$H = H_0 + H_{e-e} + H_{e-ph} + H_{e-pt}, \quad (13)$$

Each term has the following expressions:

$$H_0 = \sum_i \epsilon_i a_i^\dagger a_i + \sum_{v\vec{q}} \hbar \Omega_{v\vec{q}} b_{v\vec{q}}^\dagger b_{v\vec{q}} + \sum_\mu \hbar \omega_\mu c_\mu^\dagger c_\mu, \quad (14)$$

$$H_{e-e} = \frac{1}{2} \sum_{ijkl} v_{kl}^{ij} a_i^\dagger a_j^\dagger a_k a_l, \quad (15)$$

$$H_{e-ph} = \sum_{ijv\vec{q}} G_{ij}^{v\vec{q}} a_i^\dagger a_j (b_{v\vec{q}} + b_{v-\vec{q}}^\dagger), \quad (16)$$

$$H_{e-pt} = i\hbar \sum_{ij\mu} (g_{ij}^\mu a_i^\dagger a_j c_\mu - g_{ij}^{\mu*} a_j^\dagger a_i c_\mu^\dagger) - \frac{i\hbar e_0}{m_0} \sum_{ij} \vec{M}_{ij} \cdot \vec{A}(t) a_i^\dagger a_j, \quad (17)$$

where $a_i^\dagger$, $b_{v\vec{q}}^\dagger$, $c_\mu^\dagger$ are the electron, phonon, and photon creation operators, $\epsilon_i$ is the single-particle energy for electron, $v_{ijkl}$ is the Coulomb matrix element, $\vec{M}_{ij}$ is the optical matrix element, $\vec{A}$ is the vector potential of the external pump light, $\Omega_{v\vec{q}}$ is the frequency of phonon, $G_{ij}^{v\vec{q}}$ ($g_{ij}^\mu$) is the electron-phonon (electron-photon) coupling matrix element, and $\omega_\mu$ is the frequency of photon[9].

*6.3 Optical Bloch equation*

We study the carrier dynamics using the optical Bloch equations (OBE). With the

Heisenberg equation of motion, the time evolution of the electron population can be described by[8]:

$$\begin{aligned} \frac{d\rho_l}{dt} &= 2Re(\sum_i \Omega_{il} p_{il}), \\ \frac{dp_{if}}{dt} &= (i\omega_{if} - \gamma)p_{if} + \Omega_{fi}(\rho_f - \rho_i), \end{aligned} \quad (18)$$

where $\rho_l(t) = \langle a_l^\dagger a_l \rangle(t)$ is the electron population at a pseudo-Landau level with quantum index $l$, $p_{if}$ is the microscopic polarization, $\Omega_{fi}$ is the Rabi frequency, and $\omega_{if} = \frac{E_f - E_i}{\hbar}$ is the frequency difference between initial and final states. The change of the electron populations for the probe energy (i.e., $\Delta\rho_f - \Delta\rho_i$) as a function of time is directly proportional to the measured reflection change ($\Delta R/R$) presented in our study[10]. The electron-electron Coulomb scattering is the main factor that dominates the non-equilibrium carrier dynamics and is considered for the many-particle interactions in the OBE. The electron-electron Coulomb interaction is written as[8]:

$$H_{e-e} = \frac{1}{2} \sum_{ijkl} v_{ij}^{kl} a_i^\dagger a_j^\dagger a_k a_l. \quad (19)$$

The many-particle scattering rate by the electron-electron interactions can be written by:

$$\Gamma_f^{cc,in}(t) = \frac{2\pi}{\hbar} \sum_{abc} V_{bc}^{fa} (V_{fa}^{bc} - V_{fa}^{cb})(1 - \rho_a)\rho_b \rho_c L_\gamma(\Delta E_{bc}^{fa}), \quad (20)$$

$$\Gamma_i^{cc,out}(t) = \frac{2\pi}{\hbar} \sum_{abc} V_{bc}^{ia} (V_{ia}^{bc} - V_{ia}^{cb})\rho_a (1 - \rho_b)(1 - \rho_c) L_\gamma(\Delta E_{bc}^{ia}), \quad (21)$$

where $V_{bc}^{ia}$ is the Coulomb matrix element. It can be written as:

$$V_{bc}^{ia} = \sum_{\vec{q}} V(\vec{q}) \rho_{ib}(\vec{q}) \rho_{ac}(-\vec{q}) = \alpha_{n_b n_c}^{n_i n_a} \delta_{\xi_i,\xi_b} \delta_{\xi_a,\xi_c} c_{bc}^{ia} \frac{e_0}{\epsilon_0} \int d\vec{q} \frac{1}{\vec{q}} \tilde{F}_b^i(\vec{q}) \tilde{F}_c^a(\vec{q}), \quad (22)$$

where $\tilde{F}_b^i(\vec{q})$ is the combined form factor, which considers the shape of different Landau levels:

$$\tilde{F}_b^i(\vec{q}) = sgn(n_i n_b) F_{n_b-1,m_b}^{n_i-1,m_i}(\vec{q}) + F_{n_b,m_b}^{n_i,m_i}(\vec{q}), \quad (23)$$

$F_{n_b-1,m_b}^{n_i-1,m_i}(\vec{q})$ is the general form factor[11], and the coefficient $\alpha_{n_b n_c}^{n_i n_a}$ is written as:

$$\alpha_{n_b n_c}^{n_i n_a} = (\sqrt{2})^{\delta_{n_i,0} + \delta_{n_a,0} + \delta_{n_b,0} + \delta_{n_c,0}}, \quad (24)$$

and $c_{bc}^{ia}$ is a constant. Lorentzian $L_\gamma(\Delta E_{bc}^{ia})$ for the conservation of energy is defined as:

$$L_\gamma(\Delta E) = \frac{1}{\pi} \frac{\gamma}{\Delta E^2 + \gamma^2}, \quad (25)$$

where $\Delta E_{bc}^{ia} = E_b - E_i + E_c - E_a$. Due to the presence of many electrons and the surrounding material, the Coulomb potential will be screened and renormalized as[8]:

$$v_q \to \frac{v_q}{\epsilon_r(\vec{q},\omega)\epsilon_b}. \quad (26)$$

The dielectric function $\epsilon_r(\vec{q},\omega)$ can be calculated as:

$$\epsilon_r(\vec{q},\omega) = 1 - \frac{V_q}{\epsilon_b}\Pi(\vec{q},\omega), \quad (27)$$

where $\epsilon_b$ is the background dielectric constant, $\Pi(\vec{q},\omega)$ is the polarizability calculated in the random phase approximation (RPA) according to Goerbig et al.[12] Thus, the Coulomb interaction-modified OBEs now read as[8]:

$$\begin{aligned}\frac{d\rho_l}{dt} &= \Gamma_l^{cc,in}(1-\rho_l) - \Gamma_l^{cc,out}\rho_l, \\ \frac{dp_{if}}{dt} &= \frac{1}{2}\left(\Gamma_i^{cc,in} + \Gamma_i^{cc,out} + \Gamma_f^{cc,in} + \Gamma_f^{cc,out}\right)p_{if}.\end{aligned} \quad (28)$$

We point out that the main many-body interactions considered in the rise process are electron-electron scattering, in which the photoexcited carrier distribution rapidly broadens as pairs of electrons scatter to lower and higher energy states. Since the pump energy is larger than the probe energy, we assume that the dynamical picture of the rise process is that the excited carriers are first pumped to the initial Landau level, $LL_i$, and then inject to a lower energy level that is the final Landau level, $LL_f$, through electron-electron scattering channels. The experimentally measured probe signals indicate the time of carrier filling in the target $LL_f$ level. In Fig. 4, the calculated probe rise times for the target $LL_f$ level show a clear slow-down of carrier relaxation when the pseudo-magnetic field intensity is increased from 20 to 80 tesla (4 cases: 20 T, 40 T, 60 T, 80 T). Based on the pump (1030 nm) and probe energies (1450 nm), the target levels are chosen differently for different pseudo-magnetic fields. In principle, the dominated electron-electron scattering requires the energy and momentum conservation, but the non-equidistant energetic separation feature of the pseudo-Landau levels in strained graphene can effectively suppress the carrier scattering because energy conservation is not fulfilled[13]. This is in fact in contrast to the data for unstrained graphene having a short rise time of a few tens of femtoseconds owing to the existence of ample electron-electron scattering channels within the continuous Dirac bands. A larger pseudo-magnetic field and a larger separation of pseudo-Landau levels lead to a longer time for carriers to relax from the pump energy level to the probe energy level, which is highly consistent with our experimental observation (Fig. 3).

**Supplementary Figures**

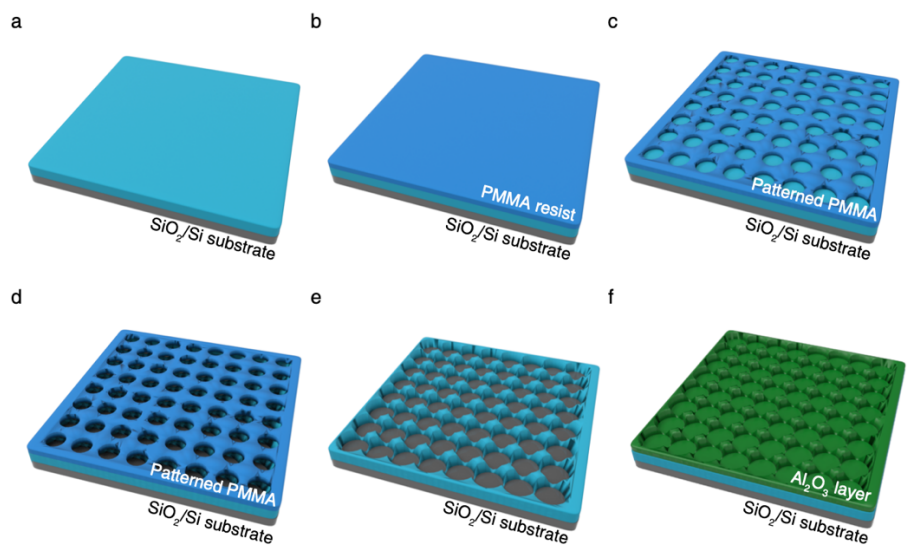

**Supplementary Figure 1 | Fabrication process of a nanostructured substrate. a-b,** The PMMA resist was covered on a 300-nm-thick $SiO_2$/Si substrate. **c,** Patterned PMMA is used as an etch mask. **d,** The sample was then soaked in BOE for 3 min 30 sec. **e-f,** The PMMA layer was removed using acetone, IPA, DI water, and $O_2$ plasma, followed by 20-nm-thick $Al_2O_3$ deposition using ALD.

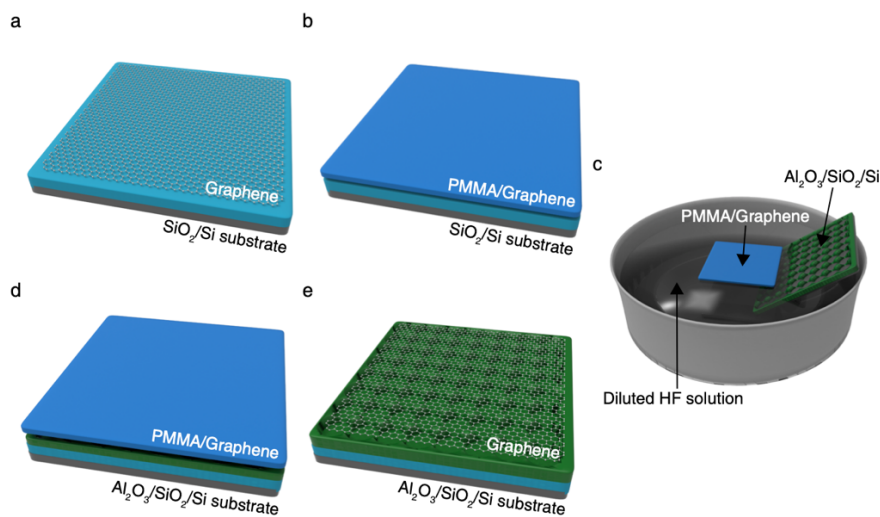

**Supplementary Figure 2 | Graphene wet transfer to the nanostructured substrate. a-b,** The PMMA supporting layer was covered on a graphene/$SiO_2$/Si sample. **c,** The PMMA/graphene layer was floated in a diluted HF solution to etch way the underlying $SiO_2$ layer. The PMMA/graphene layer was then transferred to DI water. The nanostructured

substrate was used to fish the PMMA/graphene layer. **d,** The sample was then dried with standing at an angle to make a tight adhesion between graphene and the substrate by capillary force. **e,** PMMA was removed using acetone, IPA and DI water.

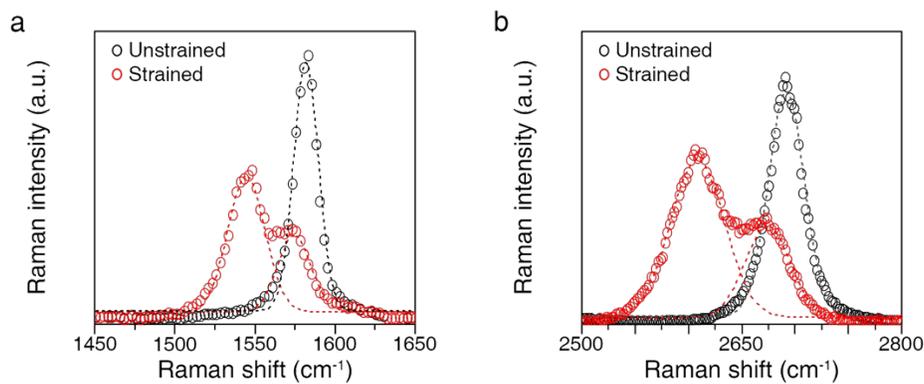

**Supplementary Figure 3 | Raman analysis of unstrained and strained graphene a,** Raman spectra of G mode for unstrained (black) and strained (red) graphene. **b,** Raman spectra of 2D mode for unstrained (black) and strained (red) graphene. By applying high strain on graphene using nanopillars, 2D and G peaks split into $2D^+$ / $2D^-$ and $G^+$ / $G^-$, respectively. Corresponding 2D and G peaks have been fitted with two Lorentzian peaks in each spectrum (dashed line).

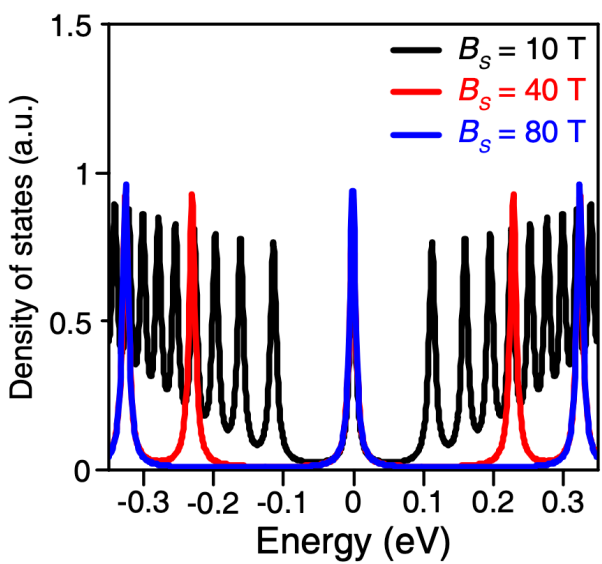

**Supplementary Figure 4 |** Calculated LDOS in strained graphene with 10 T (black), 40 T (red), and 80 T (blue) of pseudo-magnetic fields.

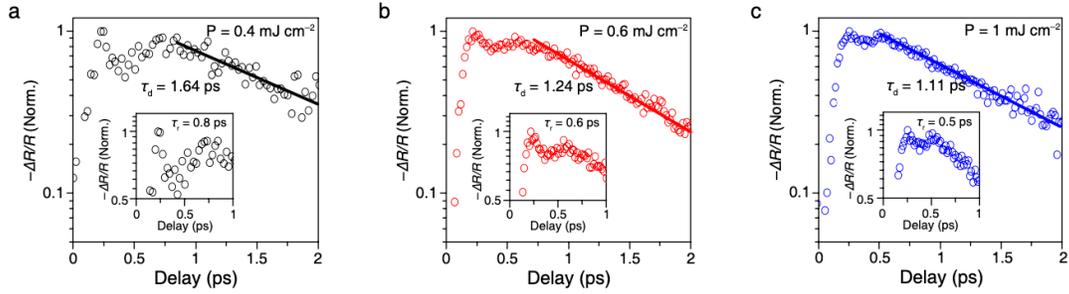

**Supplementary Figure 5 | Pump-dependent carrier dynamics of the nanopillar device.** Normalized reflection changes of the nanopillar sample at different pump fluence of **a,** 0.4 mJ cm$^{-2}$, **b,** 0.6 mJ cm$^{-2}$, and **c,** 1 mJ cm$^{-2}$. All measurements were performed at 4 K. Symbols are measurement data; lines are fitting data for decay regions. Inset: Normalized reflection changes of the nanopillar sample from 0 to 1 ps. Both rise and decay times were reduced by increasing pump fluence.